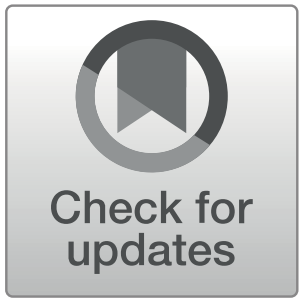

# Reversible denoising and lifting based color component transformation for lossless image compression

Roman Starosolski [1]

## Abstract

An undesirable side effect of reversible color space transformation, which consists of lifting steps (LSs), is that while removing correlation it contaminates transformed components with noise from other components. Noise affects particularly adversely the compression ratios of lossless compression algorithms. To remove correlation without increasing noise, a reversible denoising and lifting step (RDLS) was proposed that integrates denoising filters into LS. Applying RDLS to color space transformation results in a new image component transformation that is perfectly reversible despite involving the inherently irreversible denoising; the first application of such a transformation is presented in this paper. For the JPEG-LS, JPEG 2000, and JPEG XR standard algorithms in lossless mode, the application of RDLS to the RDgDb color space transformation with simple denoising filters is especially effective for images in the native optical resolution of acquisition devices. It results in improving compression ratios of all those images in cases when unmodified color space transformation either improves or worsens ratios compared with the untransformed image. The average improvement is 5.0–6.0% for two out of the three sets of such images, whereas average ratios of images from standard test-sets are improved by up to 2.2%. For the efficient image-adaptive determination of filters for RDLS, a couple of fast entropy-based estimators of compression effects that may be used independently of the actual compression algorithm are investigated and an immediate filter selection method based on the detector precision characteristic model driven by image acquisition parameters is introduced.

✉ Roman Starosolski
rstarosolski@polsl.pl; rstaros@gmail.com

[1] Institute of Informatics, Silesian University of Technology, Akademicka 16, 44-100 Gliwice, Poland

## 1 Introduction

An undesirable side effect of reversible color space transformations employed by lossless image compression algorithms to reduce correlation among image components is that they contaminate transformed components with noise from other components. Reducing the correlation of components improves compression ratios, but the noise increase may reduce or thwart the expected transformation effects. Reversible color space transformations are constructed using Lifting Steps (LSs) because LSs have many practical advantages, like in-place computation or perfect reversibility [5], but it is these steps that propagate noise between components.

To solve the above-mentioned problem, replacing LSs with reversible denoising and lifting steps (RDLSs) was proposed. RDLS prevents noise propagation by exploiting denoising filters while retaining other effects and advantages of LS. It was first applied to the simple RDgDb transformation [34], and then to three other more complex color space transformations [36]; the method of replacing LSs with RDLSs is general and may be used with any transformation exploiting LSs, for instance with the Discrete Wavelet Transformation (DWT) [35]. A transformation modified by applying RDLS (RDLS-modified transformation) retains the desirable effect of the original transformation (e.g., the correlation reduction) but avoids the unwanted side effect of noise propagation. The perfect reversibility of the RDLS-modified transformation is preserved despite the inherently lossy nature of denoising, which may be surprising. For this reason, the first application of RDLS is described in detail as a background in this paper and the forward and inverse transformation is illustrated with a step-by-step example. In addition, to facilitate understanding of the method, the notation for the transformed components is changed compared with other publications on RDLS.

By an image-adaptive denoising filter selection, the RDLS-modified transformation may be better-suited to characteristics of the processed data. Filters might be selected by checking the actual compression ratio obtained by a compression algorithm for the image transformed using RDLSs exploiting different denoising filters. From a practical standpoint, however, the complexity of such a selection method would be too high. Here, research is reported in which, based on the transformation effects observed for real images contaminated to different degrees with noise, interesting properties of RDLS were identified that allowed proposing alternative filter selection methods. The first filter selection method exploits an entropy-based estimator of compression effects that is applied directly to the transformed image components, or to components subjected to prediction (linear or nonlinear). This estimator is significantly faster than compression algorithms and is independent of the actual compression algorithm because the optimal filter selection depends more on the images than on the compression algorithm. Another filter selection method proposed in this study consists of building a model of an image acquisition device, which links the image acquisition parameters with suitable denoising filters. This method, for some types of images, allows for a virtually costless selection of filters (single table lookup using image acquisition parameters) and obtains good results in comparison to the above-mentioned estimators.

The research was done for RDLS-modified RDgDb (RDLS-RDgDb) using several sets of test images (including standard sets as well as ones prepared especially for this research) and three significantly different standard compression algorithms (JPEG-LS [42], JPEG 2000 [41], and JPEG XR [7]). The application of RDLS to the RDgDb color space transformation with simple denoising filters was found especially effective for images in the native optical resolution of acquisition devices. It results in improving the compression ratios of all these images and for all compression algorithms in cases when unmodified color space transformation either improves or worsens ratios compared with the untransformed image. The average improvement is 5.0–6.0% for two out of the three sets of such images, whereas average ratios of images from standard test-sets are improved by up to 2.2%.

The remainder of this paper is organized as follows. In subsections of Section 2, after presenting methods of constructing color space transformations for image compression (Sect. 2.1) and characterizing LS (Sect. 2.2), RDLS is presented (Sect. 2.3) along with its application to RDgDb (Sect 2.4) and a detailed example of forward and inverse RDLS-RDgDb (Sect. 2.5). The experimental procedure is described in Section 3. The effects of RDLS-modified transformation on the entropy and bitrates of transformed image components are evaluated and discussed in Section 4.1. In Section 4.2, from a practical standpoint, a fast and compression algorithm-independent entropy-based estimator for selecting denoising filters for RDLS is investigated. Then, an immediate estimator based on image acquisition parameters is proposed in Section 4.3. Section 5 summarizes the research.

## 2 Background

### 2.1 Color space transformations for image compression

It is known that the red, green, and blue primary color components of the RGB color space are highly correlated for natural images [20, 24]. A common approach to RGB color image compression is the independent compression of the image components obtained using a transformation from RGB to a less correlated color space. This technique is employed by standards including JPEG 2000 [16, 41], DICOM incorporating JPEG 2000 [21], JPEG-LS (part II of the standard) [11], and JPEG XR [7, 14].

For a specific image, by using Principal Component Analysis (PCA), the image-dependent Karhunen–Loève transformation (KLT) may be obtained that optimally decorrelates the image [24]. Note that the PCA/KLT is much better known for its use as a spatial transformation than for the transformation of color space. There are two important reasons why other color space transformations are used instead of it. The optimal decorrelation does not lead to the best lossless compression ratios [34] and, from a practical standpoint, PCA/KLT requires too much time to be computed each time an image gets compressed. Therefore, instead of directly using KLT, fixed transformations are constructed by performing PCA on a representative image set. It is then assumed that the obtained KLT transformation matches individual images both within and outside the set. Many such fixed transformations have been constructed and practically exploited. Established transformations, like the YCgCo from JPEG XR, were constructed for typical continuous-tone photographic images (YCgCo was constructed using the Kodak set described in Section 3 [20]). Such transformations are only effective as long as the characteristic of the image to be transformed matches the assumed characteristic.

Images whose characteristic is different may require either color space transformations constructed especially for them or transformations that are image-adaptive. Two increasingly often exploited and diverse classes of such images are worth mentioning here: screen content images and medical images. Screen content images may be computer-generated (ray-traced, drawn, etc.), composed from others (including natural ones), or be screenshots. There currently is a growing interest in compression of screen content images [22], the recent standard of video and image coding HEVC [17] includes an extension for these images [43]. The volume of medical images requiring storage is ever-growing, also the employed resolutions and depths tend to increase and these images often require long-term storage. Fortunately, PCA/KLT is not the only approach allowing adaptation of the color space transformation to a given image. A simple method named Adaptive Color Transform is used in the screen content coding extension of HEVC [43], where either the image block data is encoded directly in the RGB color space or the YCgCo transformation is applied before coding if it improves the coding effects. In [38, 39] it was shown that for a given image or a region of an image, an adaptive selection of a transformation from a large family of simple transformations may be done at the cost of a slight increase in color image transformation process complexity. An adaptive method based on Singular Value Decomposition is significantly more complex, yet it is simpler than computing KLT for the image [31].

Certain images require lossless compression. Medical images are a typical example because lossy compression of various medical image modalities used for diagnostic purposes may be forbidden by law [4, 19]. The lossy image compression should not be employed when processing (e.g., storing, transmitting) images of unknown purpose. The color space transformation for lossless compression has to be reversible when transformed components are stored using integers (i.e., it has to be integer-reversible). Such transformation is usually based on an irreversible transformation and is obtained using a lifting scheme [5, 40] for factorization of the transformation matrix into a sequence of LSs. The necessary and sufficient condition for factorization is that the transformation matrix determinant is 1 or −1 [9]. See Eq. 1 for an example matrix of a forward reversible component transformation from JPEG 2000 (RCT) and Eq. 2 for one of the forms of the actual lifting-based RCT (both forward and inverse):

$$\begin{bmatrix} Y \\ Cu \\ Cv \end{bmatrix} \approx \begin{bmatrix} 1/4 & 1/2 & 1/4 \\ 0 & -1 & 1 \\ 1 & -1 & 0 \end{bmatrix} \begin{bmatrix} R \\ G \\ B \end{bmatrix}, \tag{1}$$

$$\begin{array}{l} Cu = B - G \\ Cv = R - G \\ Y = G + \lfloor (Cu + Cv)/4 \rfloor \end{array} \quad \Leftrightarrow \quad \begin{array}{l} G = Y - \lfloor (Cu + Cv)/4 \rfloor \\ R = Cv + G \\ B = Cu + G \end{array}, \tag{2}$$

where the floor symbol $\lfloor x \rfloor$ denotes the greatest integer not exceeding $x$. This matrix, without additional assumptions (e.g., without specifying how to obtain the integer value of the $Y$ component), is a close approximation of the actual transformation only.

In [34], a couple of simple transformations were proposed as a result of departing from the traditional method of constructing transformations for lossless image compression based on a transformation for lossy compression, which in turn is based on PCA/KLT for a specific image set. They resulted in better compression ratios than both the established transformations (RCT, YCoCg-R [14]) and PCA/KLT. The simplest, RDgDb, is presented in Eq. 3 with its inverse transformation. Both require only two integer subtractions per pixel:

$$\begin{array}{l} Db = G - B \\ Dg = R - G \\ R = R \end{array} \quad \Leftrightarrow \quad \begin{array}{l} R = R \\ G = R - Dg \\ B = G - Db \end{array} \tag{3}$$

RDgDb was found to be the most universal because it resulted in the best average bitrates for two out of the three compression algorithms tested (JPEG 2000, JPEG XR) and the second-best bitrate for the third (JPEG-LS [12, 42]). The above transformation was independently proposed and investigated in [39] as part of a large family of reversible transformations. However, for images used in that research, its performance evaluated with a different methodology was not the best.

An unwanted side effect of established color space transformations, like RCT, is that while removing correlation they contaminate all transformed components with noise from other components. Components of images acquired by consumer acquisition devices operating at high photon flux are mainly affected by additive white Gaussian noise [2]; however, noise parameters may differ among components due to differences among primary color detector cells or detector filters. The luminance component of RCT ($Y$) contains a fraction of noise from all the primary color components, and each chrominance component ($Cu$ and $Cv$) contains noise from two primary colors. The key difference between RCT and RDgDb is in replacing the luminance component by the red primary color component. Now only the chrominance components may contain more noise than the primary color components. This may explain the good performance of the simpler transformation that indeed decorrelates images worse than established ones (the average correlation of RDgDb transformed image components was the second largest of the transformations examined in [34]). However, bitrates of chrominance components usually are significantly better than bitrates of individual primary color components despite an increased amount of noise. The method presented in this study is aimed at constructing a reversible transformation that would remove the correlation but without increasing the amount of noise.

### 2.2 Lifting step

LS exploited by a color space transformation is a simple operation on a pixel's components that modifies one of the components by adding to it a linear combination (a weighted sum) of some or all of the other components of the same pixel; the sum may be negated. A practical advantage of employing LSs is that the transformation may be computed in place, i.e., the transformed components overwrite their untransformed counterparts and, therefore, do not require any additional memory space. A typical reversible color space transformation is constructed as a sequence of LSs and, because of the in-place operation, is implemented as such a sequence. However, alternative methods of implementation are possible—for instance, RCT components might be obtained by directly implementing formulas from Eq. 2 and treating untransformed components as read-only data. In RCT implemented using LSs, the $Cu$ component, after computing it as $Cu = B - G$, is stored in place of the $B$ component, overwriting it. In other words, the $B$ component is modified by subtracting $G$ from it; after this modification, $B$ is renamed as $Cu$. Next, the modified $R$ becomes $Cv$, and then the modified $G$ becomes $Y$. For further information on LSs, the lifting scheme, and their use in color space transformations the Reader is referred to [5, 9, 20, 40].

To introduce RDLS (see the next subsection), LS is generalized here as:

$$C_x \leftarrow C_x \oplus \mathrm{f}(C_1, \ \dots \ , C_{x-1}, C_{x+1}, \ \dots \ , C_n) \tag{4}$$

where $C_i$ is the $i$-th component of a pixel, $C_x$ is the component being modified by the step, and $n$ is the number of color space components. Any components may be used as arguments of function f, except for $C_x$. Demonstration of the in-place operation of LS is an advantage of the notation used in Eq. 4. Another advantage is that it allows seeing clearly that noise may get propagated by LS only to $C_x$ (because no other component gets modified by this LS) and thus only from other components used in this step, i.e., from arguments of the function f. The step is reversible provided that the function f is deterministic and that the operation $\oplus$ is reversible, i.e., there is an inverse operation $\otimes$ such that $c = a \oplus b \Leftrightarrow a = c \otimes b$. For color space transformations, the operation $\oplus$ is a subtraction (or a negated subtraction), but for binary planes, for example, the exclusive or could be used instead.

### 2.3 Reversible denoising and lifting step

To avoid the propagation of noise from other components used while performing LS to the component $C_x$ that is modified by the step, denoising filters are integrated into the step. RDLS (Eq. 5) is constructed based on LS (Eq. 4) by applying denoising to the arguments of function f:

$$C_x \leftarrow C_x \oplus \mathrm{f}\left(C_1^d, \ \ldots \ , C_{x-1}^d, C_{x+1}^d, \ \ldots \ , C_n^d\right) \tag{5}$$

where $C_i^d$ is the denoised $i$-th component of a pixel obtained using a denoising filter. Denoising is not an in-place operation; computing $C_i^d$ and using it as an argument for function f does not alter $C_i$. In other words, we temporarily denoise all components from which noise could be propagated to the component $C_x$ being modified by the step. The only difference between LS (Eq. 4) and RDLS (Eq. 5) is in the denoising applied to arguments of function f, but this difference has significant consequences.

LS-based color space transformation may be performed for each pixel independently of other pixels. The RDLS sequence, constructed on the basis of a color space transformation, is a transformation of the entire image components, not of the color space because denoising of a pixel's component usually requires access to the same component of neighboring pixels. Now, let us recall that in Eq. 4 we could use any components as arguments of function f, except for $C_x$. In Eq. 5, we denoise these arguments, thus in denoising them we may use any component of any pixel, except for $C_x$ of the pixel to which RDLS is being applied because otherwise this $C_x$ would be indirectly used by f. For RDLS, not only the function f has to be deterministic but also all the denoising filters must be deterministic.

In a general case, denoising (i.e., computing the temporarily denoised components to be used as arguments of function f) cannot be performed before or after a transformation that is a sequence of RDLSs but has to be done within it because the same component may be used in more than one step of the transformation. Denoising may be applied to both the untransformed and the already transformed component. In consequence, different filters may be used not only for different components but also for the same component in different transformation steps. Furthermore, because the filter should match the characteristic of the data to be denoised, filters for each RDLS should be selected in an image-adaptive way.

Despite the inherently lossy nature of denoising, RDLS exploiting denoising and the transformation consisting of several such steps are perfectly and easily invertible. An inverse of the transformation consisting of RDLSs is obtained by applying inverses of RDLSs:

$$C_x \leftarrow C_x \otimes \mathrm{f}\left(C_1^d, \ \ldots \ , C_{x-1}^d, C_{x+1}^d, \ \ldots \ , C_n^d\right) \tag{6}$$

in an order opposite to the one employed by the forward transformation. In the next subsection, forward and inverse RDLS-RDgDb are presented, and an example of their application is presented in Section 2.5. Naturally, the same denoising filters must be used for the same components in inverse RDLSs as were applied during forward RDLSs.

Actual denoising filters are not perfect because they remove noise only partially. Consequently, RDLS transfers a fraction of noise from the component being denoised to the transformed component. On the other hand, they may introduce distortions, e.g., blurring, to the component being denoised, making it less suitable for removing correlation. Thus, finding the right denoising filter is crucial because the introduced distortions might affect the bitrate of the transformed component more than the noise reduction, resulting in bitrate worsening.

RDLS was proposed for transformations for noisy images, but it appeared more general. Two interesting special filters for RDLS were found effective in improving the compression ratios in practice: the None filter for which $C_i^d = C_i$ [35] and the Null filter, for which $C_i^d = 0$ [36]. The None filter turns RDLS into a regular LS if it is applied to all arguments of function f (see Eqs. 4 and 5). Thanks to the None filter, the original transformation becomes a special case of the RDLS-modified transformation. The Null filter, if applied to all arguments of function f, may result in effective skipping of the step (the steps of color space transformations and DWT are by the Null filter turned into either $C_x \leftarrow C_x$ or $C_x \leftarrow -C_x$). These filters may turn the RDLS-modified transformation into a lifting transformation or cause skipping it as a whole or partially—because different filters may be selected for different steps by the filter selection method employed for a specific image. The Null filter provides an additional mechanism of adaptation of the RDLS-modified transformation in the case when overall transformation effects (both unwanted side effects and desirable ones) result in worsening of the compression ratio. Thanks to this filter, RDLS may be effective also for the noise-free data.

The method of replacing LSs with RDLSs is of a general nature and is applicable to other transformations that are sequences of LSs. RDLS has already been found effective for color space transformations more complex than RDgDb (RCT, YCoCg-R, and LDgEb) [36]. Significant bitrate improvements were obtained for grayscale nonphotographic images by the application of RDLS to a multiple-level multidimensional DWT in JPEG 2000 lossless compression [35, 37].

### 2.4 RDLS-modified RDgDb transformation

For the practical application of RDLS, the RDgDb color space transformation (Eq. 3) was selected because of its good performance in lossless compression and simplicity. To apply RDLS to RDgDb, first, let us rewrite RDgDb using the notation as in Eqs. 4–6 and thus obtaining the following Eq. 7:

$$\begin{array}{lll} step\,1: & C_3 \leftarrow -C_3 + C_2 & \\ step\,2: & C_2 \leftarrow -C_2 + C_1 & \Leftrightarrow \\ step\,3: & C_1 \leftarrow C_1 & \end{array} \begin{array}{ll} step\,1: & C_1 \leftarrow C_1 \\ step\,2: & C_2 \leftarrow -C_2 + C_1 \\ step\,3: & C_3 \leftarrow -C_3 + C_2 \end{array} \tag{7}$$

This notation shows explicitly how the in-place computations are performed and uses the same symbol for a pixel's component before and after modifying its value by the step. For this reason, the $C_1$, $C_2$, and $C_3$ components have dual meaning compared with the traditional notation used in Eqs. 2 and 3—they denote $R$, $G$, and $B$ components of the untransformed image, respectively, and $R$, $Dg$, and $Db$ components of the transformed image, respectively. In a general case, the steps must be performed in a specified order. The $R = R$ formula in Eq. 3 (and its equivalent $C_1 \leftarrow C_1$ in Eq. 7) may be seen as just additional information that $R$ is a component of the transformed pixel, or as a special case of LS, where $a \oplus b = a \otimes b = a$ (recall the reversible operations $\oplus$ and $\otimes$ defined in Section 2.2). In the other steps, $a \oplus b = a \otimes b = -a + b$.

Next, each LS (Eq. 4) of the RDgDb transformation (Eq. 7) should be simply replaced with RDLS (Eq. 5) constructed based on it by applying denoising filters to arguments of the function f, i.e., to all components used in the step except the one that is modified by it. The obtained RDLS-RDgDb transformation is presented in Eq. 8.

$$\begin{array}{ll} step\,1: & C_3 \leftarrow -C_3 + C_2^d \\ step\,2: & C_2 \leftarrow -C_2 + C_1^d \\ step\,3: & C_1 \leftarrow C_1 \end{array} \Leftrightarrow \begin{array}{ll} step\,1: & C_1 \leftarrow C_1 \\ step\,2: & C_2 \leftarrow -C_2 + C_1^d \\ step\,3: & C_3 \leftarrow -C_3 + C_2^d \end{array} \tag{8}$$

$C_1$, $C_2$, and $C_3$ denote $R$, $G$, and $B$ components of the untransformed image, respectively, and $R$, $dDg$, and $dDb$ components of the transformed image, respectively. To distinguish between components of RDLS-RDgDb and RDgDb, by $dDg$ and $dDb$ denoted are RDLS-RDgDb components obtained using RDLSs constructed based on LSs of RDgDb resulting in obtaining $Dg$ and $Db$ components, respectively.

The third steps of the forward RDgDb and RDLS-RDgDb, as well as their inverses (the first steps of the inverse transformations), do not require any computations. They are presented in Eqs. 7 and 8 for the sole purpose of illustrating the application of RDLS to a sample lifting-based transformation.

The lifting-based color space transformation of an image may be performed in a pixel-by-pixel regime, for which all transformation steps are applied to each pixel in turn, or step-by-step, meaning a lifting step is applied to all pixels before the next step is applied. These regimes are equivalent. For the RDLS-modified transformation, the filter depends on the regime. A typical denoising filter applied to a component of a pixel does not exploit other components and calculates the denoised component based on the same component of pixels inside a window whose center is the pixel being denoised. Naturally, if we process an image pixel-by-pixel, then all components of all the already processed pixels are transformed. Assuming the raster scan order of processing, except for the top left and bottom right corner, the filter window contains both transformed and untransformed pixels. Thus, a sophisticated denoising filter should be employed that is either able to take advantage of both transformed and untransformed components or uses only a part of the window that contains either only transformed or only untransformed pixels. The step-by-step regime is in practice simpler to apply. In this regime, for each image component, except for the component being modified in the current step, either all pixels are transformed or all are untransformed. In the remainder of this study, the step-by-step regime is employed.

## 2.5 RDLS-RDgDb example

Let us transform a noisy $4 \times 4$ pixel image using the RDLS-RDgDb transformation (Eq. 8). The top row of panels in Fig. 1 (panels a–c) shows the untransformed $C_1$, $C_2$, and $C_3$ components of the image (i.e., $R$, $G$, and $B$, respectively). For denoising, we use a simple low-pass linear averaging filter with a $3 \times 3$ pixel window. The filtered pixel component is

| 64 | 94 | 56 | 72 |
|---|---|---|---|
| 66 | 71 | 50 | 98 |
| 81 | 79 | 77 | 91 |
| 68 | 66 | 73 | 54 |

(a) Untransformed $C_1$ ($R$)

| 72 | 72 | 99 | 76 |
|---|---|---|---|
| 99 | 97 | 61 | 82 |
| 95 | 56 | 73 | 92 |
| 66 | 59 | 91 | 97 |

(b) Untransformed $C_2$ ($G$)

| 62 | 98 | 67 | 77 |
|---|---|---|---|
| 66 | 75 | 75 | 51 |
| 58 | 79 | 95 | 64 |
| 91 | 94 | 69 | 94 |

(c) Untransformed $C_3$ ($B$)

| 85 | 83 | 81 | 80 |
|---|---|---|---|
| 82 | 80 | 79 | 81 |
| 79 | 77 | 79 | 83 |
| 69 | 73 | 78 | 88 |

(d) $C_2^d$ (denoised untransformed $C_2$)

| 23 | -15 | 14 | 3 |
|---|---|---|---|
| 16 | 5 | 4 | 30 |
| 21 | -2 | -16 | 19 |
| -22 | -21 | 9 | -6 |

(e) Transformed $C_3$ ($dDb$)

| 74 | 67 | 74 | 69 |
|---|---|---|---|
| 76 | 71 | 76 | 74 |
| 72 | 70 | 73 | 74 |
| 74 | 74 | 73 | 74 |

(f) $C_1^d$ (denoised untransformed $C_1$)

| 2 | -5 | -25 | -7 |
|---|---|---|---|
| -23 | -26 | 15 | -8 |
| -23 | 14 | 0 | -18 |
| 8 | 15 | -18 | -23 |

(g) Transformed $C_2$ ($dDg$)

**Fig. 1** RDLS-RDgDb transformation of a sample 4 × 4 pixel image. Presented are components of the untransformed image (a–c), temporarily denoised components $C_2^d$ and $C_1^d$ used during the transformation (d, f), and transformed components $C_3$ and $C_2$ (e, g); the transformed $C_1$ is not shown because $C_1$ is not affected by RDLS-RDgDb

calculated as an average of components from the window and rounded to the nearest integer; the number of pixels used is smaller than the window size at the image edges.

### 2.5.1 Forward transformation

We apply step 1 of the forward RDLS-RDgDb transformation (from Eq. 8) to the pixel surrounded in Fig. 1 a–c by a solid line. Step 1 ($C_3 \leftarrow -C_3 + C_2^d$) modifies the $C_3$ component of the pixel by assigning to it a difference between the $C_2^d$ (the denoised $C_2$) and the intensity of the $C_3$ component before applying step 1. We calculate the $C_2^d$ as the average of component intensities within the area surrounded in Fig. 1b by a dashed line (in this case the average is 80). Therefore, the value of the transformed $C_3$ component, whose intensity before the transformation was 75, is calculated as $-C_3 + C_2^d = -75 + 80 = 5$. In the same way, we apply step 1 to all the other image pixels; Fig. 1d presents $C_2^d$ of all pixels and Fig. 1e presents the $C_3$ component after applying step 1 to all pixels. While applying step 1 to a certain pixel, the denoised $C_2$ component $C_2^d$ is needed only temporarily; denoising may be performed "on the fly" because $C_2^d$ is used only once and only in this step, so there is no need to store it. After applying step 1 to all pixels, the image consists of the transformed $C_3$ component (i.e., the *dDb* component of RDLS-RDgDb transformed image, Fig. 1e) and the untransformed $C_1$ and $C_2$ components (Fig. 1 a and b, respectively); the untransformed $C_3$, as well as the denoised $C_2^d$, are not available in further steps.

Similarly, we apply step 2 ($C_2 \leftarrow -C_2 + C_1^d$), which modifies the $C_2$ component. For each pixel, the transformed $C_2$ (i.e., *dDg* of the transformed image, Fig. 1g) is calculated as the difference between the denoised $C_1$ ($C_1^d$, Fig. 1f) and the untransformed $C_2$ (Fig. 1b). Step 3 ($C_1 \leftarrow C_1$) does not require any computations, so we may skip it. The RDLS-RDgDb transformed image thus consists of components *R*, *dDg*, and *dDb* (Fig. 1 a, g, and e, respectively).

### 2.5.2 Inverse transformation

During the inverse RDLS-RDgDb, we apply the inverses of the forward transformation steps in reverse order. The $C_1$, $C_2$, and $C_3$ components of the transformed image are *R*, *dDg*, and *dDb*, respectively. We skip step 1, which is an inverse of the skipped step 3 of the forward transformation. Then, we apply step 2 of the inverse RDLS-RDgDb ($C_2 \leftarrow -C_2 + C_1^d$, the inverse of the forward transformation step 2:$C_2 \leftarrow -C_2 + C_1^d$), which modifies the $C_2$ component. For each pixel, the inverse-transformed $C_2$ equals the difference between the denoised $C_1$ and the transformed $C_2$. Because denoising is performed using exactly the same data as in step 2 of the forward transformation (i.e., the same $C_1$ shown in Fig. 1a), it results in the same denoised component $C_1^d$ as in the forward transformation (Fig. 1f), which in turn allows the perfect reconstruction of the untransformed $C_2$ component of the original image, i.e., the *G* component (Fig. 1b), despite the inherently lossy nature of the denoising employed. After applying step 2 to all pixels, the image consists of the untransformed $C_1$ and $C_2$ components (i.e., *R* and *G*, respectively) and the transformed $C_3$ component (*dDb*).

Finally, for all pixels, we apply step 3 ($C_3 \leftarrow -C_3 + C_2^d$, the inverse of the forward step 1:$C_3 \leftarrow -C_3 + C_2^d$), which modifies the $C_3$ component. Denoising is based on the same data as in step 1 of the forward transformation (i.e., on the already untransformed $C_2$, Fig. 1b), so we obtain the same denoised $C_2^d$component (Fig. 1d) and reconstruct the untransformed $C_3$ (*B*, Fig. 1c).

## 3 Experimental procedure and implementations

The effects of RDLS were evaluated by comparing bitrates of lossless image compression algorithms as well as the entropy of components of RDgDb (Eqs. 3 and 7) and RDLS-RDgDb (Eq. 8). Besides the Null and None special case filters described in Section 2.3, simple Smoothing filters were used, i.e., low-pass linear averaging filters with a 3 × 3 pixel window. The filtered component intensity was calculated using the Smoothing filter as a weighted arithmetic mean of intensities of the same component of pixels from the window. At the edges of the images, the number of pixels used was smaller than the window size. Eleven Smoothing filters of different denoising strengths were employed. The denoising strength was controlled by setting a weight *w* of the center point in the weighting mask (in the range 1 to 1024; integer powers of 2 only), while the weights of neighboring points were fixed to 1; therefore, greater *w* results in weaker denoising.

Three significantly different standard image compression algorithms in the lossless mode were used in experiments: JPEG-LS, JPEG 2000, and JPEG XR. JPEG-LS is a standard of ISO/IEC and ITU-T for primarily lossless compression of still images which originates from the LOCO-I algorithm [12, 42]. The baseline standard defines a low-complexity predictive image compression algorithm with entropy coding using the modified Golomb–Rice code family. JPEG 2000 is an ISO/IEC and ITU-T standard lossy and lossless image compression algorithm based on DWT image decomposition and arithmetic coding [16, 41]. JPEG XR is an ISO/IEC and ITU-T lossy and lossless standard algorithm that is designed primarily for high-quality, high dynamic range photographic images and is based on Discrete Cosine Transformation image decomposition and adaptive Huffman coding [7, 14]. The SPMG/UBC JPEG-LS implementation, version 2.2 [30], JasPer implementation of JPEG 2000 by Adams, version 1.900 [18], and a standard reference implementation of JPEG XR, version 1.6 [15] were used.

The compression ratio or bitrate $r$, expressed in bits per pixel (bpp), is calculated as $r = 8e/s$, where $s$ is the number of pixels in the image component, $e$ is the total size in bytes of the compressed component, including compressed file format headers; hence, smaller $r$ means better compression. The memoryless entropy of the component $H_0$, also expressed in bpp, is calculated as:

$$H_0 = -\sum_{i=0}^{N-1} p_i \log_2 p_i \tag{9}$$

where $N$ is the alphabet size (256 in the case of an 8-bit color component, 512 for chrominance component) and $p_i$ is the probability of occurrence of intensity $i$ in the component.

To determine quickly the right filter for denoising an image component, we may simply use the transformed component entropy as an estimator of component compression effects. However, because image compression algorithms exploit the fact that nearby pixels are correlated, the effects of transformation on component compressibility may be better estimated by the entropy of the component residual image, which can be obtained by subtracting from each pixel the predicted pixel value. Two predictors were investigated: AVG, which is a simple linear predictor predicting that the pixel intensity equals the average of its upper and left neighbor; and MED, which is a simple nonlinear edge-detecting predictor used, among others, in the JPEG-LS algorithm [42]. When the acquisition device characteristic and parameters of the acquisition process are known, then the denoising filter parameters may be estimated directly from the image acquisition process parameters—e.g., see [1, 2]. Inspired by the above approach, an acquisition process model for images of which the exposure parameters are available (sets A3 and A3-red.3 described below) is proposed in this study. This model determines the denoising filters based on the image resolution and ISO speed used for image acquisition; it is described in detail in Section 4.3.

The evaluation was performed for the sets of 8-bit RGB test images shown below. Besides the standard image sets, three other sets containing images in the optical resolution of the acquisition device were used because it could be expected that a simple denoising filter would give better results on high-resolution unprocessed images (i.e., images that were not subjected to processing like resizing, noise removal, sharpening, etc.). Images subjected to size reduction only were also added.

- Waterloo – a set of eight color images from the University of Waterloo, Fractal Coding and Analysis Group repository, used for a long time in image processing research (Fig. 2 a–c). The set contains eight natural photographic and artificial images, including the well-known “lena” and “peppers.” Image sizes vary from 512 × 512 to 1118 × 1105, available from [26].
- Kodak – a set of 23 photographic images released by the Kodak Corporation. The set is frequently used in color image compression research. All images are of size 768 × 512, available from [27].

- EPFL – a set of 10 high-resolution images used at the École Polytechnique Fédérale de Lausanne for subjective JPEG XR quality evaluation [6]. Image sizes range from 1280 × 1506 to 1280 × 1600, available from [28].
- A1 – these images (Fig. 2 d–f) were originally acquired from a 36 mm high-quality diapositive film using a Minolta Dimage 5400 film scanner. After size reduction and conversion to grayscale, they were used in research on grayscale image compression algorithms [32]. Set A1 contains three images in the device's optical resolution. Image sizes vary from 7376 × 4832 to 7424 × 4864, available from [29].
- A2 – a set of 17 images acquired from negatives using a general-purpose Agfa e50 scanner (see Fig. 2g–i) in the device's optical resolution. Image sizes range from 1620 × 1128 to 1740 × 1164, available from [29].
- A3 – a set of 116 images acquired using an Olympus Stylus XZ-2 consumer camera. Because the camera detector uses a Bayer-pattern RGGB color filter array, the grayscale Bayer-pattern image was converted to a color image by taking for each group of RGGB subpixels the average of two green subpixels as the intensity of the green pixel component and the intensities of the red and blue subpixels as the intensities of the red and blue pixel components, respectively. Thus, the image resolution is as close to the detector resolution as possible without interpolation of all components (i.e., equal in the case of red and blue components, lower for the green component). The total number of pixels in the color image equals ¼ of the total number of subpixels in the camera detector. For extracting Bayer-pattern images from RAW camera files, dcraw version 9.24 [25] was used. To allow analyzing how noise affects the RDLS-modified transformation, the A3 set also contains six series of images (Fig. 2 j–o). Each series is taken from the same scene reaching the detector (fixed manual focus and aperture were applied) but captured with different ISO speeds (100 to 12,800 in eight steps) resulting in different amounts of noise. These series are not perfect; due to the wind, the movement of the Sun, etc., there are small differences between the images in a given series. The A3 images are noticeably sharper than the A1 and A2 images. Image sizes are 1992 × 1493, available from [29].
- Sets A1-red.3, A2-red.3, and A3-red.3, all of whose images were reduced three times from A1, A2, and A3, respectively; a pixel of a reduced-size image is calculated as an average of nine pixels from a full-size image.

Free tools used for resizing, color component transformations, prediction, etc. are available from [8]. A free research implementation for examining the effects of RDLS-RDgDb and other transformations on the Reader's images was prepared [10].

## 4 Results and discussion

### 4.1 Effects of RDLS on entropy and bitrates of transformed chrominance components

This subsection describes the preliminary research that allowed determination of some useful RDLS properties and finding effective methods for estimating transformation effects. In this part of the research, only the Smoothing filter was used and, because RDgDb and RDLS-

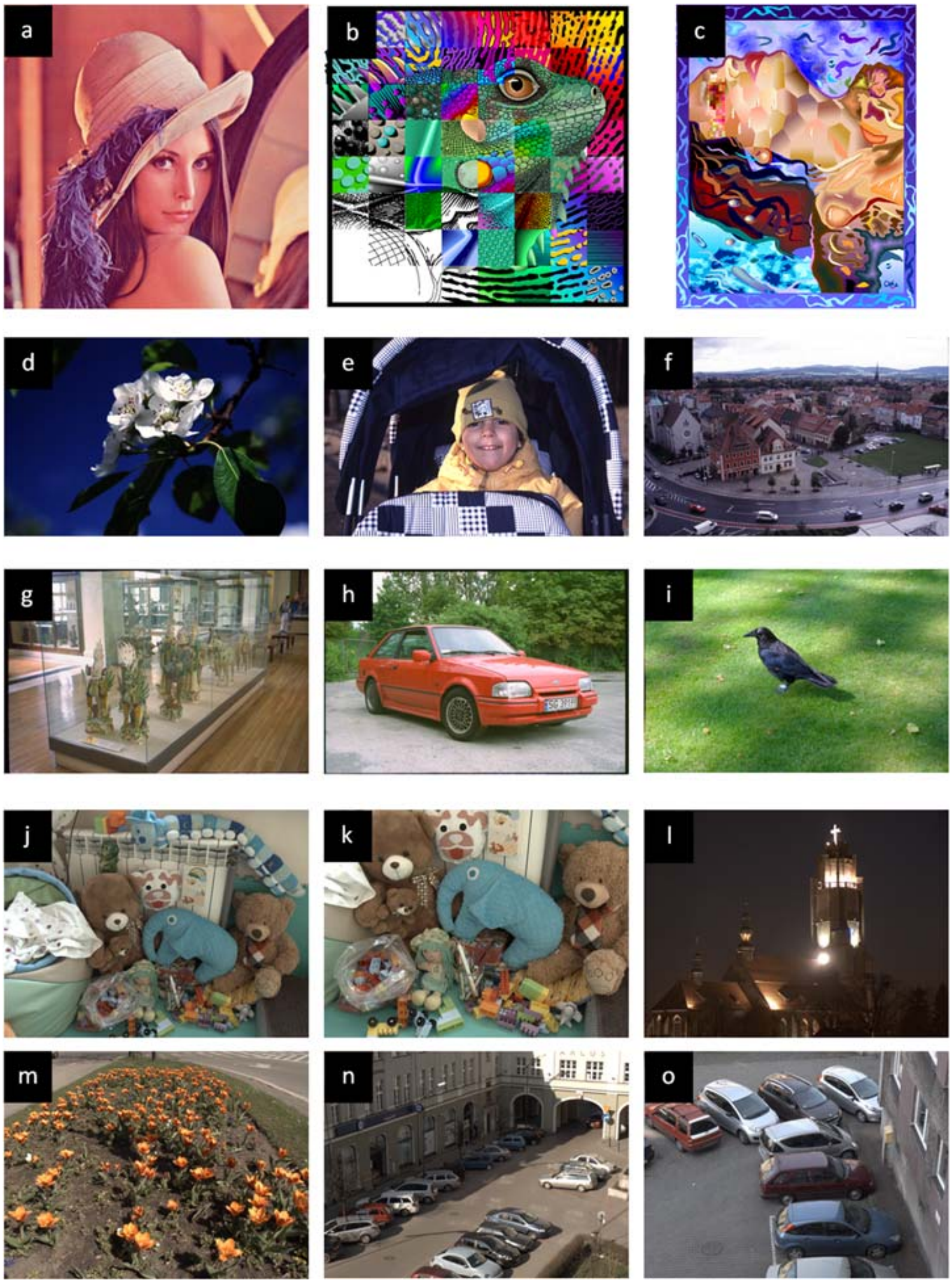

**Fig. 2** Sample images used in the research. Images "lena" (a), "frymire" (b), and "serrano" (c) from the Waterloo set, the A1 set (d–f), three images from the A2 set (g–i), and images from each series contained in the A3 set (j–o)

RDgDb affect only two of the three image components (that are transformed into the chrominance components *Dg*/*dDg* and *Db*/*dDb*), RDLS effects were analyzed only for these components.

The average entropies of RDgDb chrominance components, for each of nine sets of test images, and average entropy changes after applying RDLS are presented in Table 1. The best denoising filter (i.e., Smoothing filter) was selected for each image and image component individually. The results are also presented for residual RDLS-RDgDb chrominance component images obtained using AVG and MED predictors. The effects of RDLS are more pronounced for the entropy of the residual component images, the most pronounced are for MED (however, similar conclusions may be drawn even in the case of RDLS-RDgDb chrominance components). Looking at the entropy changes for residual images obtained using the MED predictor, we see that entropy is increased for Waterloo and Kodak images only (negligibly for the latter), which suggests that the application of RDLS only with the simple Smoothing filter to Waterloo images may worsen bitrates. The opposite may be expected for more image sets, especially for A2 and A3 images due to an entropy reduction of the *Dg* component of over 13%.

The effects of applying RDLS on RDgDb chrominance component bitrates for lossless JPEG-LS, JPEG 2000, and JPEG XR algorithms are presented in Tables 2, 3, and 4, respectively. Average RDLS-RDgDb bitrate changes, as compared with RDgDb, are reported for the best denoising filter selected for each image individually (in columns labeled $\Delta r_{\mathrm{bitrate}}$) and for the denoising filter that results in the best average bitrate for a given set (column $\Delta r_{\mathrm{set}}$). Denoising filters were selected independently for each algorithm and independently for *dDg* and *dDb*. The Smoothing filter selected for the set is also reported. We can see that for a given set and component, either the same or adjacent Smoothing filter central point weight is the best one for all algorithms, indicating that a denoising filter should be selected for the given image set rather than for the compression algorithm. However, selecting the average best filter for a set produces bitrates worse than those that may be obtained by selecting the filter for each image individually. The bitrates are worse by up to approximately 0.3% for images in acquisition device resolution and up to approximately 3% for others. Therefore, we hereafter focus on the results of selecting the denoising filter for each image and component individually.

A noticeable bitrate increase due to RDLS is observed for Waterloo images and the JPEG-LS algorithm—over 2.5% for both chrominance components. Actually, this change is a result of the much greater bitrate deterioration in the case of the two computer-generated images containing many sharp edges. The bitrates of “frymire” (Fig. 2b) and “serrano” (Fig. 2c) were worsened by over 22% and 10%, respectively. Predictive JPEG-LS outperforms other algorithms for artificial images. The Smoothing filter changes the characteristics of such images. Histograms of these images may be globally or locally highly sparse [23, 33] and application of the Smoothing filter results in an increase of the number of active levels. For other algorithms, the Waterloo bitrates are worsened by no more than 0.4%. Kodak bitrates are worsened by no more than 0.06%, and for all the other sets the RDLS application results in bitrate improvement.

For the A2 and A3 sets, RDLS resulted in an improvement of the *Dg* component bitrate for all algorithms of over 12%. Results for the former set were obtained using the strongest denoising tested ($w = 1$) for each individual image, suggesting that even greater improvements could be possible if even stronger denoising was used. For the *Db* component, the bitrate improvement is smaller (2.1–2.9%), but for most A2 images, the strongest denoising was the best. In general, RDLS effects on bitrates are similar to the observed effects on residual image entropy, especially to those obtained using MED. For A1, the improvements are smaller (2.7–5.0% for *Dg*, below 1% for *Db*), yet still practically useful. Compared with full-size images, bitrates of reduced-size images are less improved by RDLS, e.g., *Dg* bitrates of A2-red.3x and A3-red.3x are improved by 4.6–6.9%.

**Table 1** Effects of RDLS on the entropy of *Dg* and *Db* components

| Set | *Dg/dDg* | | | | | | *Db/dDb* | | | | | |
|---|---|---|---|---|---|---|---|---|---|---|---|---|
| | component | | predicted AVG | | predicted MED | | component | | predicted AVG | | predicted MED | |
| | $H_0$ | $\Delta H_0$ | $H_0$ | $\Delta H_0$ | $H_0$ | $\Delta H_0$ | $H_0$ | $\Delta H_0$ | $H_0$ | $\Delta H_0$ | $H_0$ | $\Delta H_0$ |
| Waterloo | 6.4489 | 1.15% | 3.5465 | 1.17% | 2.7440 | 2.68% | 6.3846 | 1.25% | 3.5554 | 1.34% | 2.7679 | 2.77% |
| Kodak | 5.6283 | 0.01% | 2.8993 | 0.03% | 2.7112 | 0.05% | 5.6355 | 0.01% | 2.9571 | −0.04% | 2.8793 | 0.01% |
| EPFL | 6.3105 | 0.02% | 3.3350 | −0.23% | 3.2630 | −0.17% | 6.2441 | 0.00% | 3.5963 | −0.92% | 3.5175 | −1.46% |
| A1 | 5.0821 | −0.11% | 2.2484 | −4.49% | 2.2080 | −5.12% | 6.2161 | −0.01% | 2.6813 | −0.16% | 2.5769 | −0.24% |
| A2 | 6.2611 | −1.84% | 4.9566 | −12.16% | 5.1361 | −13.01% | 6.7713 | −0.33% | 5.4503 | −2.66% | 5.6381 | −2.84% |
| A3 | 6.0461 | −3.97% | 5.2656 | −14.14% | 5.4417 | −15.35% | 5.9153 | −0.75% | 5.0694 | −2.39% | 5.2540 | −2.45% |
| A1-red.3x | 5.0475 | 0.00% | 2.7309 | −1.84% | 2.7930 | −2.49% | 6.1965 | −0.02% | 3.3058 | −0.14% | 3.3418 | −0.11% |
| A2-red.3x | 6.0348 | −0.65% | 4.4527 | −4.43% | 4.5144 | −5.24% | 6.5321 | −0.06% | 4.9038 | −0.85% | 4.9917 | −1.05% |
| A3-red.3x | 5.5786 | −1.22% | 4.3292 | −5.93% | 4.4119 | −6.83% | 5.5091 | −0.38% | 4.1134 | −2.23% | 4.1936 | −2.09% |

Entropies reported for components and residual component images obtained using AVG and MED predictors. $H_0$ – memoryless entropy of component after unmodified RDgDb transformation (bpp), $\Delta H_0$ – entropy change due to applying RDLS with denoising filter selected for the image and component

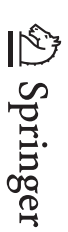

**Table 2** Effects of RDLS on JPEG-LS bitrates of *Dg* and *Db* components

| Set | *Dg/dDg* | | | | *Db/dDb* | | | |
|---|---|---|---|---|---|---|---|---|
| | $r$ | $\Delta r_{\text{bitrate}}$ | $w_{\text{set}}$ | $\Delta r_{\text{set}}$ | $r$ | $\Delta r_{\text{bitrate}}$ | $w_{\text{set}}$ | $\Delta r_{\text{set}}$ |
| Waterloo | 2.7110 | 2.62% | 1024 | 3.02% | 2.7807 | 2.94% | 1024 | 3.14% |
| Kodak | 2.5276 | 0.06% | 1024 | 0.06% | 2.6981 | 0.03% | 1024 | 0.03% |
| EPFL | 3.0266 | −0.14% | 1024 | 0.03% | 3.2570 | −1.80% | 1024 | 0.03% |
| A1 | 2.0804 | −5.03% | 4 | −4.94% | 2.4208 | −0.51% | 8 | −0.44% |
| A2 | 4.9253 | −12.83% | 1 | −12.83% | 5.4428 | −2.94% | 1 | −2.93% |
| A3 | 5.2158 | −15.28% | 2 | −14.95% | 5.0314 | −2.25% | 8 | −1.96% |
| A1-red.3x | 2.6001 | −2.32% | 16 | −2.05% | 3.1462 | −0.10% | 64 | −0.07% |
| A2-red.3x | 4.3451 | −5.23% | 16 | −4.53% | 4.8241 | −0.99% | 32 | −0.40% |
| A3-red.3x | 4.1827 | −6.93% | 16 | −3.88% | 3.9884 | −2.00% | 64 | −1.02% |

$r$ – bitrate of component after unmodified RDgDb transformation (bpp); $\Delta r_{\text{bitrate}}$ – RDLS-RDgDb bitrate change when selecting a denoising filter for each image and component individually; $w_{\text{set}}$ – Smoothing filter center point weight resulting in the best average component bitrate for a given set; $\Delta r_{\text{set}}$ – RDLS-RDgDb bitrate change when denoising all set images using the Smoothing filter with the $w_{\text{set}}$ weight

Looking at the results of individual images (not shown in the tables), it may be noticed that for each image in the acquisition device optical resolution, as well as for all reduced-size A1 images, RDLS improved the bitrates of both chrominance components with all the compression algorithms examined. For reduced-size A2 and A3 images, the bitrates are improved on average, but not for each individual image. The opposite may be said about Waterloo images, whose bitrates were noticeably increased on average but not for each individual image, e.g., an improvement was consistently observed for the well-known “lena” image (Fig. 2a).

Set A3 contains six series of images (Fig. 2 j–o) taken with ISO speed from 100 to 12,800. The images in each of these series are essentially the same noise-free images that reach the detector but are contaminated during acquisition by various amounts of noise proportional to the ISO speed. It is worth noting that ISO speeds of 3200 and above are reported by the camera as “extended” and are not used by the camera unless manually forced. Besides the series, set A3 contains many images taken with the automatic selection of exposure parameters, where usually a low ISO speed was selected. In Fig. 3 for the A3 set series, JPEG 2000 average bitrates of *R*, *G*, and *B* primary color components, *Dg* and *Db* chrominance components after RDgDb transformation, and components *dDg* and *dDb* obtained by RDLS-RDgDb are

**Table 3** Effects of RDLS on JPEG 2000 bitrates of *Dg* and *Db* components

| Set | *Dg/dDg* | | | | *Db/dDb* | | | |
|---|---|---|---|---|---|---|---|---|
| | $r$ | $\Delta r_{\text{bitrate}}$ | $w_{\text{set}}$ | $\Delta r_{\text{set}}$ | $r$ | $\Delta r_{\text{bitrate}}$ | $w_{\text{set}}$ | $\Delta r_{\text{set}}$ |
| Waterloo | 3.4148 | 0.39% | 1024 | 0.64% | 3.4718 | 0.30% | 1024 | 0.68% |
| Kodak | 2.4668 | 0.06% | 1024 | 0.06% | 2.5583 | 0.04% | 1024 | 0.04% |
| EPFL | 3.1521 | −0.18% | 512 | −0.01% | 3.3800 | −1.98% | 1024 | −0.01% |
| A1 | 2.1119 | −4.09% | 4 | −3.98% | 2.3961 | −0.71% | 4 | −0.65% |
| A2 | 4.9789 | −12.69% | 1 | −12.69% | 5.5086 | −2.70% | 1 | −2.68% |
| A3 | 5.2546 | −14.67% | 2 | −14.36% | 5.0651 | −2.29% | 8 | −2.00% |
| A1-red.3x | 2.6654 | −1.96% | 16 | −1.73% | 3.1760 | −0.10% | 64 | −0.07% |
| A2-red.3x | 4.4528 | −5.22% | 16 | −4.55% | 4.9384 | −1.01% | 32 | −0.39% |
| A3-red.3x | 4.2694 | −6.59% | 16 | −3.71% | 4.0820 | −2.24% | 64 | −1.21% |

$r$, $\Delta r_{\text{bitrate}}$, $w_{\text{set}}$, $\Delta r_{\text{set}}$ – see the Table 2 notes

**Table 4** Effects of RDLS on JPEG XR bitrates of *Dg* and *Db* components

| Set | *Dg/dDg* | | | | *Db/dDb* | | | |
|---|---|---|---|---|---|---|---|---|
| | $r$ | $\Delta r_{\text{bitrate}}$ | $w_{\text{set}}$ | $\Delta r_{\text{set}}$ | $r$ | $\Delta r_{\text{bitrate}}$ | $w_{\text{set}}$ | $\Delta r_{\text{set}}$ |
| Waterloo | 4.1733 | 0.26% | 1024 | 0.59% | 4.2199 | 0.00% | 1024 | 0.66% |
| Kodak | 3.0526 | 0.02% | 1024 | 0.03% | 3.1215 | 0.00% | 1024 | 0.01% |
| EPFL | 3.4972 | −0.15% | 512 | −0.02% | 3.7092 | −1.17% | 512 | −0.03% |
| A1 | 2.6630 | −2.77% | 4 | −2.72% | 2.9376 | −0.32% | 8 | −0.27% |
| A2 | 5.1057 | −12.04% | 1 | −12.04% | 5.6047 | −2.62% | 1 | −2.60% |
| A3 | 5.4194 | −13.87% | 2 | −13.58% | 5.2392 | −2.15% | 8 | −1.89% |
| A1-red.3x | 3.0258 | −1.46% | 16 | −1.31% | 3.4711 | −0.09% | 64 | −0.06% |
| A2-red.3x | 4.6751 | −4.65% | 16 | −4.09% | 5.1376 | −0.90% | 32 | −0.28% |
| A3-red.3x | 4.5679 | −5.93% | 16 | −3.26% | 4.3779 | −1.93% | 64 | −1.04% |

$r$, $\Delta r_{\text{bitrate}}$, $w_{\text{set}}$, $\Delta r_{\text{set}}$ – see the Table 2 notes

presented. For easier comparisons, the left-hand panel contains results for *Dg*, *dDg*, and the color components used to compute them, whereas the right-hand panel shows the *Db*, *dDb*, and the color components needed to compute *Db* and *dDb*. Only results averaged over all series are presented because individual series results are similar to the average. Figure 4 presents the average results for reduced-size images from the same series taken from the A3-red.3x set.

In both figures, we see that bitrates of the *G* component are significantly lower than bitrates of the remaining primary color components. The acquisition detector noise is reduced in the case of the *G* component during the conversion from the Bayer-pattern color filter array detector image to the color image. In the A3 set, the *G* pixel component was calculated as the average of two green subpixels from the camera detector, while the remaining color components were taken directly from single detector subpixels. Lower noise levels may also be a result of the different characteristics of the band-pass green filter and other filters constituting the color filter array. A3-red3x images are much less noisy; here also *G* contains the least noise

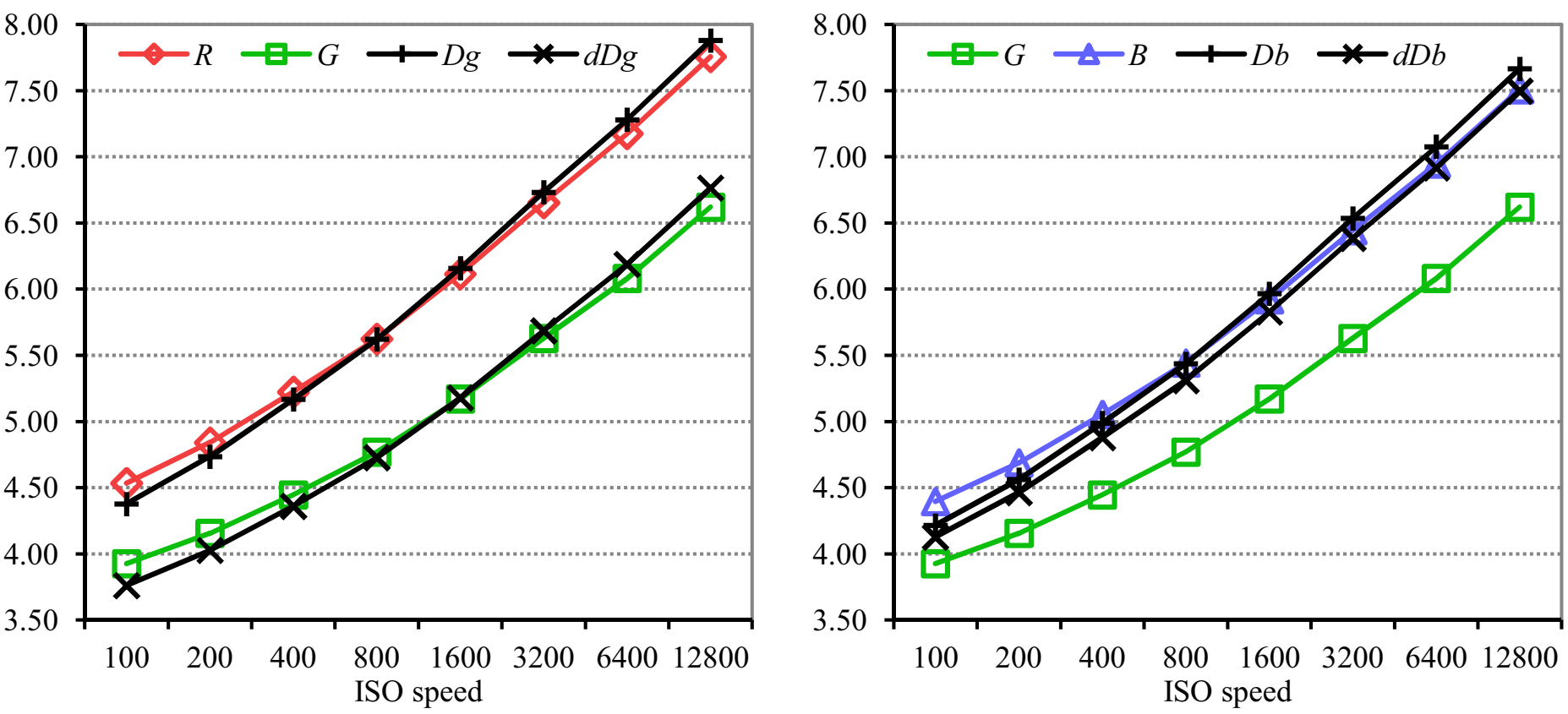

**Fig. 3** JPEG 2000 average bitrates for series of acquisition device resolution images with varying amounts of noise. JPEG 2000 average component bitrates (bpp) for six series of images of the same objects from the A3 set (Fig. 2 j–o) photographed using different ISO speeds and containing various amounts of noise proportional to the ISO speed. The left panel shows *Dg* chrominance component, its RDLS-modified variant *dDg*, and primary color components *R* and *G* used to compute them. The right panel shows the *Db* component, its RDLS variant *dDb*, and primary color components *G* and *B*

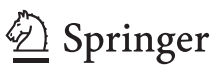

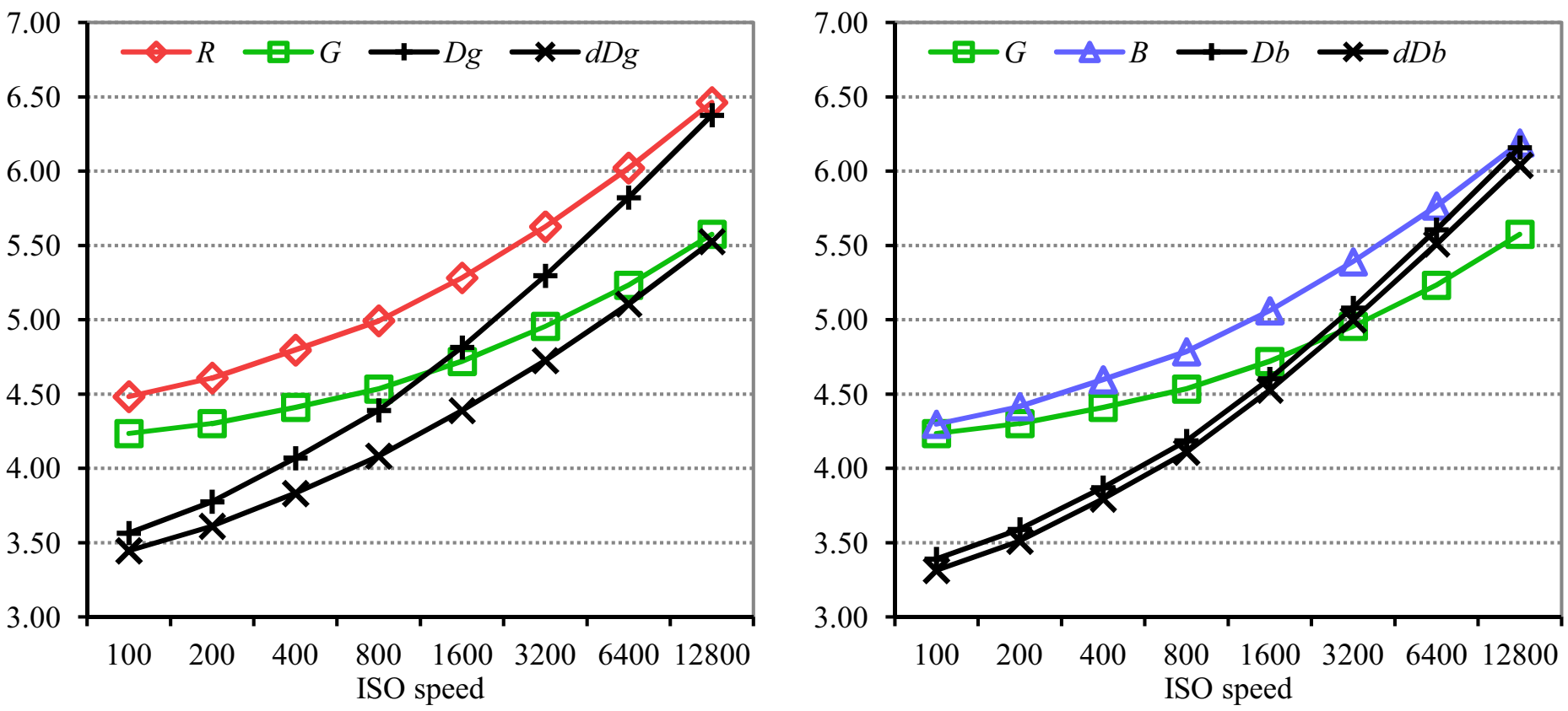

**Fig. 4** JPEG 2000 average bitrates for series of reduced resolution images with varying amounts of noise. JPEG 2000 average component bitrates (bpp) for six series of reduced-size (A3-red.3x set) images of the same objects photographed using different ISO speeds and containing various amounts of noise, proportional to the ISO speed of full-size images. The left panel shows the *Dg* chrominance component, its RDLS-modified variant *dDg*, and primary color components *R* and *G* used to compute them. The right panel shows the *Db* component, its RDLS variant *dDb*, and components *G* and *B*

because it is actually an average of 18 green subpixels, whereas other components are averages of 9 subpixels each.

As already noted, RDLS improves bitrates of chrominance components of each A3 image as well as for almost every image of the A3-red.3x set. Figures 3 and 4 show how this improvement depends on the image noise level and component. When the level of noise is high (for A3 and noisier A3-red.3x images), the bitrate of the chrominance component (i.e., *Dg* or *Db*) is close to the bitrate of the primary color component used to calculate the chrominance, whose bitrate is worse (to *R* and *B*, respectively). For the highest noise levels (noisier A3 images), the chrominance bitrate is even worse than the inferior bitrates of the primary color components. Clearly, RDgDb is not the best color space transformation for these images. When we apply RDLS, the bitrate of the RDLS-RDgDb chrominance component (*dDg* or *dDb*) is close to the bitrate of the untransformed primary color component that becomes this chrominance component during the transformation (*G* or *B*, respectively). Therefore, the RDLS bitrate improvement is high if the component being modified during transformation is compressed better than the other component used to calculate the chrominance (calculation of *Dg* in this case). For the A3 set series, the bitrate decreased from 14.1% to 15.9%. If the component being modified is compressed worse (calculation of *Db*), then both unaltered and RDLS-modified chrominance bitrates are close to the bitrate of the component being modified. We get a smaller but still practically useful bitrate decrease due to RDLS (2.1–2.3% for A3 series), which should probably be attributed to transferring less noise during transformation while still removing the correlation.

When the level of noise is low (less noisy A3-red.3x images), the effect of removing the correlation has a greater impact on the resulting bitrate than the effect of transferring noise to the chrominance component from color components, and the bitrate of chrominance is significantly lower than bitrates of both color components. An interesting field of future research is the theoretical analysis of the aforementioned RDLS behavior.

RDLS may certainly be seen as a method of improving the bitrate of chrominance components of certain types of images (e.g., images in acquisition device optical resolution that were not subject to further processing and, to a lesser extent, those processed only by size reduction), both in cases where unmodified color space transformation improves or worsens chrominance bitrates. However, for other images (Waterloo), a noticeable bitrate increase due to the RDLS application was observed. For the noisiest A3 images (see Fig. 3), although RDLS significantly improves the bitrate of the RDgDb transformed image, it is even better to compress the untransformed RGB image. The above results were obtained by RDLS with only the Smoothing filter. Using a more diverse filter set should avoid the above bitrate increase, e.g., using the Null filter that causes the RDLS-modified transformation effectively to be skipped.

## 4.2 RDLS effects from a practical standpoint

To evaluate the effects of RDLS, the entropy and bitrates of chrominance components were analyzed in the previous section, but in practice, we are more interested in the overall bitrate improvement achievable for an entire three-component image at an acceptable cost. Therefore, hereafter let us focus on bitrate improvements of three-component images due to RDLS. Based on observations from Section 4.1, selecting the special filter cases None and Null should also be allowed, besides selecting Smoothing filters. Performing actual compression of transformed components (obtained using RDLSs with different denoising filters) to determine the best filter for a specific image component may be unacceptable from a practical standpoint because it increases the compression time too much. Therefore, a practical method that allows for quickly finding the denoising filters for an image without employing an actual compression algorithm is needed.

Recalling that in the average case the optimal denoising filters for the RDLS-modified transformation are almost the same for all the compression algorithms, we check whether in order to select the denoising filter for RDLS the effects of RDLS-RDgDb on bitrate may be estimated before actual compression and independently of the compression algorithm by testing the entropy of the components of transformed images subjected to prediction. In Table 5, the bitrate improvements obtained for three-component color images by application $r_{\text{RDgDb}}$ – bitrate of an RDgDb transformed color image calculated as the sum of bitrates of three image components (bpp), $\Delta r_{\text{bitrate}}$ – bitrate change due to RDLS with the denoising filter selection (from 11 Smoothing filters, Null, and None) based on the actual compressed component bitrate, $\Delta r_{H0_\text{pMED}}$ – bitrate change for the filter selection based on the entropy of the residual component image obtained using the MED predictor of RDLS to RDgDb are reported; the effects of the denoising filter selection based on the actual bitrate obtained for a given component by the compression algorithm (shown in columns labeled $\Delta r_{\text{bitrate}}$) are compared with the effects of the selection based on the memoryless entropy of the residual component image obtained using the MED predictor (columns labeled $\Delta r_{H0_\text{pMED}}$).

**Table 5** RDLS effects for three-component images and efficiency of entropy-based estimation for RDLS filter selection

| Set | JPEG-LS | | | JPEG 2000 | | | JPEG XR | | |
|---|---|---|---|---|---|---|---|---|---|
| | $r_{\text{RDgDb}}$ | $\Delta r_{\text{bitrate}}$ | $\Delta r_{H0_\text{pMED}}$ | $r_{\text{RDgDb}}$ | $\Delta r_{\text{bitrate}}$ | $\Delta r_{H0_\text{pMED}}$ | $r_{\text{RDgDb}}$ | $\Delta r_{\text{bitrate}}$ | $\Delta r_{H0_\text{pMED}}$ |
| Waterloo | 8.8653 | −0.68% | −0.65% | 11.1428 | −0.82% | −0.81% | 13.2978 | −0.36% | −0.30% |
| Kodak | 9.5673 | 0.00% | 0.00% | 9.4754 | 0.00% | 0.00% | 10.8725 | 0.00% | 0.00% |
| EPFL | 10.3383 | −2.19% | −2.19% | 10.7240 | −1.44% | −1.44% | 11.6723 | −0.81% | −0.80% |
| A1 | 6.3869 | −3.25% | −3.20% | 6.3920 | −3.77% | −3.44% | 8.2259 | −1.03% | −0.85% |
| A2 | 14.8272 | −5.39% | −5.37% | 15.0354 | −5.22% | −5.22% | 15.4139 | −4.95% | −4.95% |
| A3 | 15.4922 | −6.01% | −5.96% | 15.6774 | −5.81% | −5.74% | 16.1822 | −5.45% | −5.39% |
| A1-red.3 | 8.3402 | −0.76% | −0.76% | 8.5181 | −0.65% | −0.65% | 9.6403 | −0.49% | −0.49% |
| A2-red.3 | 13.6966 | −2.01% | −1.99% | 14.1401 | −2.00% | −1.98% | 14.8451 | −1.78% | −1.77% |
| A3-red.3 | 13.0793 | −2.88% | −2.83% | 13.4509 | −2.81% | −2.79% | 14.3621 | −2.49% | −2.48% |

Looking at the bitrate improvements obtained without using the entropy estimation, it is clear that the greatest improvements were obtained for A2 and A3 sets; depending on the algorithm and the set, the bitrate improvement is from 5.0% to 6.0%, which is significant for the lossless image compression and useful from a practical standpoint. Sets A2 and A3 contain unprocessed images in the optical resolution of acquisition devices, the A1 set resembles them in this respect. In addition, for the A1 set, the improvements are better than in the case of standard image sets and quite good in the case of JPEG-LS and JPEG 2000 (3.1% and 3.8%, respectively). All in all, the RDLS with the simple Smoothing filter, and even simpler Null and None filters, appears an effective method for improving bitrates of images in the optical resolution of acquisition devices. Bitrates of reduced-size sets A1-red.3, A2-red.3, and A3-red.3 were improved to a smaller extent than their full-size counterparts, however, the improvement for A2-red.3 and A3-red.3 (from 1.8% to 2.8%) is still useful. Improvements for the standard sets (Waterloo, Kodak, and EPFL) do not exceed 1.0%, except for the EPFL set in the case of JPEG-LS and JPEG 2000. By using more advanced denoising filters, we would probably obtain better results for these images (as well as for images for which simple filters already result in significant bitrate improvements).

Using the entropy of residual component image obtained using the MED predictor as an estimator of the component compression effects results in overall color image bitrate improvements which are very close to those obtained when the denoising filter was selected based on the actual bitrate of the compression algorithm. They are worse by up to 0.07 percentage points for all the sets and compression algorithms except for the A1 set in the case of JPEG 2000 and JPEG XR (where the estimation effects are worse by up to 0.31 percentage points). The other estimators were tested as well: the entropy of component and entropy of residual component image obtained using the AVG predictor (not reported in Table 5), however, they were significantly worse. In addition, the correlation between the actual color image bitrate and bitrate estimated with the entropy of residual component images obtained using the MED predictor, measured with the Pearson correlation coefficient, is very high. It is at least 0.94 for all the sets and compression algorithms except for the Waterloo set in the case of JPEG 2000 and JPEG XR. This estimator is good enough to be used in practice.

The obtained results suggest that the application of RDLS to the RDgDb color space transformation with simple denoising filters is especially effective for images in the native optical resolution of acquisition devices both when the actual bitrate and its estimation using MED are employed for selecting denoising filters. Let us test the following statistical hypothesis $H_1$: for images in the native optical resolution of acquisition devices, RDLS results in mean bitrate improvement greater than for other images by at least 3 percentage points. Applying the one-tailed independent samples $t$ test (considering samples from two populations: 136 bitrate improvements due to RDLS obtained for images from sets A1, A2, and A3 and 177 improvements obtained for all the other images) at a confidence level 99.5% ($\alpha = 0.005$), we find that $H_1$ is valid for the JPEG-LS, JPEG 2000, and JPEG XR algorithms, and for both methods of selecting denoising filters.

The entropy-based estimator, although not involving actual compression and being significantly simpler than actual compression, is quite complex compared with the transformation itself. For each denoising filter, we have to denoise the color component and compute the chrominance component. For all these chrominance components, we have to compute the residual image and calculate its entropy. Overall, the denoising of components is the most time-consuming part of the entropy-estimation-based filter search process. Fortunately, it can easily be parallelized because this may be done independently for image components, separate image regions, and different filters. Another possibility is to exploit similarities among filters. Computing the Smoothing filter with 3 × 3 pixel window for a specific component of a specific pixel requires 10 integer arithmetic operations. Computing it when the center point weight is 1 requires nine operations or only five if the filter is applied to all pixels inside a contiguous area. Having computed it for a given filter window center point weight (weights of other points are 1 s), computing the filter for another center point weight can be done in just three such operations. The estimation method used in this study may be seen as an adaptation of an automatic method of selecting a color space out of a large set of color spaces presented by Strutz [39], where component transformation was selected based on the memoryless entropy of the residual component image obtained using the MED predictor. Strutz also found that computing the residual image entropy from only a small subset of image pixels (up to 10,000 pixels) is sufficient for close to optimum transformation selection; an application of a similar approach to RDLS is investigated in [36].

### 4.3 Denoising filter selection based on properties of the acquisition process

When the acquisition device characteristics and parameters of the acquisition process are known or may be determined beforehand (e.g., for a specific camera's RAW files), then the acquisition noise characteristics and proper denoising filters (or parameters of a more sophisticated denoising filter) may be estimated directly from the acquisition process parameters using the detector precision characteristics (DPC) approach (e.g., see [1, 2]). By adopting the DPC approach, a model of the acquisition process is here proposed for A3 and A3-red.3 images (because for these images the exposure parameters are available). This model assumes that noise depends on two parameters of the acquisition process: the ISO speed and the image size. The allowed sizes are: full-size (A3 images) and reduced threefold (A3-red.3 images); the allowed ISO speeds are: 100, 200, 400, 800, 1600, 3200, 6400, and 12,800. The A3 and A3-red.3 sets contain single images captured with ISO speeds 125 and 160, which are treated by the model as if they were captured with ISO speeds 100 and 200, respectively. The model, for a given image, based on parameters of this image's acquisition process selects the denoising filters (as before, for each component individually, out of 11 Smoothing filters, Null, and None). The selected filter is simply the one that was the best for images used for training the model that were of the same size and ISO speed. Such a model, once constructed for a specific acquisition process, allows immediate selection of the proper denoising filter. Table 6 presents an example model (for JPEG 2000, trained using all A3 and A3-red.3 images).

The effectiveness of the model for A3 and A3-red.3 images was tested using the leave-one-out cross-validation, i.e., for each image the model was trained using all other images (independently for each compression algorithm). Table 7 compares the effects of the denoising filter selection using the model to the selection based on the actual bitrate of compressed components. From a practical standpoint, the results are very good, especially for A3 images. For these images, using the model resulted in bitrate

**Table 6** The model of the acquisition process for determining the RDLS-RDgDb filters, trained using all A3 and A3-red.3 images for JPEG 2000 lossless compression

| ISO speed | Size: full | | Size: reduced three-fold | |
|---|---|---|---|---|
| | *dDg* | *dDb* | *dDg* | *dDb* |
| 100 | Smoothing, $w=4$ | Smoothing, $w=32$ | Smoothing, $w=16$ | Smoothing, $w=64$ |
| 200 | Smoothing, $w=2$ | Smoothing, $w=16$ | Smoothing, $w=8$ | Smoothing, $w=32$ |
| 400 | Smoothing, $w=2$ | Smoothing, $w=16$ | Smoothing, $w=8$ | Smoothing, $w=32$ |
| 800 | Smoothing, $w=2$ | Smoothing, $w=16$ | Smoothing, $w=4$ | Smoothing, $w=32$ |
| 1600 | Smoothing, $w=2$ | Smoothing, $w=8$ | Smoothing, $w=4$ | Smoothing, $w=32$ |
| 3200 | Null | Smoothing, $w=4$ | Smoothing, $w=2$ | Smoothing, $w=16$ |
| 6400 | Null | Smoothing, $w=4$ | Smoothing, $w=2$ | Smoothing, $w=16$ |
| 12,800 | Null | Smoothing, $w=2$ | Smoothing, $w=1$ | Smoothing, $w=8$ |

$w$ – Smoothing filter center point weight

improvements worse than those obtained when denoising filters were selected based on the actual bitrate of the compression algorithm by up to 0.15 percentage points only. For A3 images, the effects of entropy estimation were little better (recall Table 5 and see Fig. 5), but the entropy estimation is quite complex compared with the actual transformation, whereas the model-based filter selection is immediate—it requires a single table lookup per component. Worse effects, than in the case of A3 images, were obtained by the model for their reduced-size counterparts from the A3-red.3x set. However, the approach seems useful also for them because it allows obtaining the majority of the bitrate improvement obtainable based on testing the actual bitrates.

## 5 Conclusions and further work

An unwanted side effect of a reversible color space transformation that consists of LSs is that while removing correlation it contaminates transformed components with noise from other components. To avoid increasing noise while preserving other transformation properties, by integrating denoising into LSs a reversible image component transformation was obtained that is a sequence of RDLSs. The method was evaluated for several test-sets of images using JPEG-LS, JPEG 2000, and JPEG XR compression algorithms in lossless mode, RDgDb color space transformation, simple linear denoising filters (11 Smoothing filters), and special RDLS filter cases: Null and None.

**Table 7** Efficiency of the model of the acquisition process as an estimator of transformation effects for lossless component compression

| Set | JPEG-LS | | | JPEG 2000 | | | JPEG XR | | |
|---|---|---|---|---|---|---|---|---|---|
| | $r_{\mathrm{RDgDb}}$ | $\Delta r_{\mathrm{bitrate}}$ | $\Delta r_{\mathrm{DPC}}$ | $r_{\mathrm{RDgDb}}$ | $\Delta r_{\mathrm{bitrate}}$ | $\Delta r_{\mathrm{DPC}}$ | $r_{\mathrm{RDgDb}}$ | $\Delta r_{\mathrm{bitrate}}$ | $\Delta r_{\mathrm{DPC}}$ |
| A3 | 15.4922 | −6.01% | −5.86% | 15.6774 | −5.81% | −5.65% | 16.1822 | −5.45% | −5.33% |
| A3-red.3 | 13.0793 | −2.88% | −2.26% | 13.4509 | −2.81% | −2.22% | 14.3621 | −2.49% | −2.02% |

$r_{\mathrm{RDgDb}}$ – bitrate of an RDgDb transformed color image (bpp), $\Delta r_{\mathrm{bitrate}}$ – bitrate change due to RDLS with the denoising filter selection (from 11 Smoothing filters, Null, and None) based on the actual compressed component bitrate, $\Delta r_{\mathrm{DPC}}$ – bitrate change due to the denoising filter selection based on the model exploiting the properties of the image acquisition process

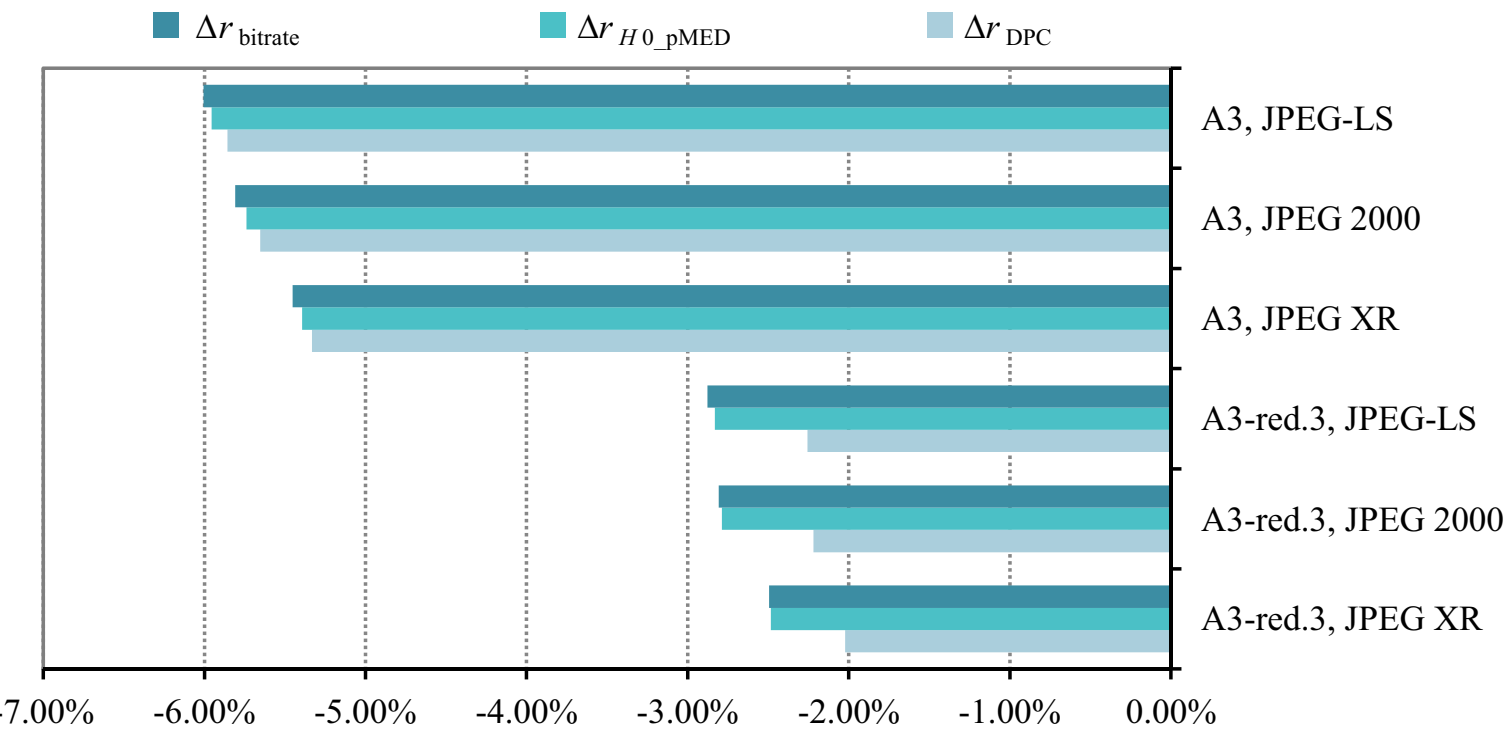

**Fig. 5** Comparison of the effects of various methods of denoising filter selection. Presented are average image bitrate changes due to RDLS with the denoising filter selection (from 11 Smoothing filters, Null, and None) based on: actual bitrate obtained using the compression algorithm ($\Delta r_{\text{bitrate}}$), estimated bitrate obtained using entropy estimator exploiting the MED predictor ($\Delta r_{H0_\text{pMED}}$), and the model exploiting properties of the image acquisition process ($\Delta r_{\text{DPC}}$)

RDLS was found effective for all algorithms for all tested unprocessed images in the optical resolution of acquisition devices. RDLS-RDgDb resulted in improving the bitrates of all such images in cases when unmodified color space transformation either improves or worsens bitrates compared with the untransformed image. Overall, the average improvement of the three-component color image bitrate was 5.0–6.0% for two out of the three sets of such images (A2 and A3). A smaller improvement was observed for the third set (1.2–3.8%) and for images of reduced sizes (0.5–2.8%). For two out of the three standard test image sets, among others containing computer-generated and sharpened images, the improvements were from 0.4% to 2.2%. The RDLS effects were found to be similar for various compression algorithms and the memoryless entropy of the residual component image obtained using the MED predictor proved to be an efficient estimator of transformation effects for the selection of denoising filters. This estimator may be applied before performing an actual compression and independently of the compression algorithm used. For images whose acquisition parameters were available (A3 and A3-red.3), a model inspired by the DPC approach was proposed that allowed for a virtually costless selection of denoising filters and was especially effective for images in the optical resolution of acquisition devices (A3).

The method presented in this paper is of a general nature and is applicable to other transformations that are sequences of LSs. RDLS has already been found effective for color space transformations that are more complex than RDgDb (RCT, YCoCg-R, and LDgEb) [36]. Significant bitrate improvements were obtained for grayscale nonphotographic images by the application of RDLS to a two-dimensional DWT in JPEG 2000 lossless compression [35, 37]. In the ongoing research, RDLS is applied to the three-dimensional DWT in JPEG 2000 standard part 10 (JP3D, [3, 13]) lossless compression of volumetric medical images. It is also employed to combine predictive and transform (DWT) coding in lossless grayscale image compression. Further research plans include joining RDLS-modified color space transformations and the RDLS-modified DWT to obtain an image compression algorithm effective for a broad range of image types. Other fields of future research are the theoretical analysis of RDLS effects on noisy image bitrates and finding better denoising filters.

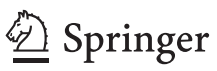

**Acknowledgments** The author thanks Tytus Bernaś, Piotr Fabian, and the anonymous reviewers for their constructive comments. This work was supported by the 02/020/BK_19/0171 grant from the Institute of Informatics, Silesian University of Technology.

## Compliance with ethical standards

**Conflict of interests** Patent PL 229208 was obtained for the transformation presented in this paper.

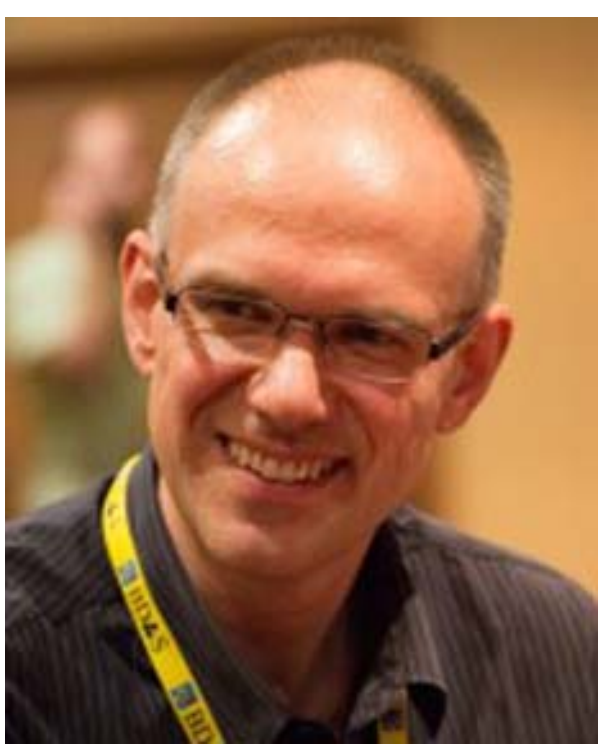

**Roman Starosolski** received his M.Sc., Ph.D., and D.Sc. degrees in computer science from the Silesian University of Technology in 1995, 2002, and 2017 respectively. Currently, he is an Associate Professor at the Silesian University of Technology in the Institute of Informatics. He is the author and co-author of more than 30 papers and has written five book chapters; he was granted four patents. His works are cited over 300 times according to Google Scholar (100 according to Web of Science). His current research interests include algorithms and standards of image and video compression, lossless image compression, image processing, biomedical imaging, lifting-based reversible transforms (color space transforms and DWT), and the recently developed method of reversible denoising and lifting steps.